\documentclass[12pt]{article}
\usepackage{amsmath}
\usepackage{amssymb}

\usepackage{graphics, graphicx}

\usepackage{caption}

\usepackage{bm}

\usepackage{color}
\usepackage[colorlinks,citecolor=blue]{hyperref}

\usepackage[superscript]{cite}

\title{Exotic localization for the bound states in the non-reciprocal two-particle Hubbard model}

\author
{Huan-Yu Wang,$^{1*}$ Ji Li,$^{2}$ Wu-Ming Liu,$^{3}$ Lin Wen, $^{4\dagger}$ Xiao-Fei Zhang$^{5\ddagger}$ \\
\normalsize{$^{1}$Fujian Key Laboratory of Quantum Information and Quantum Optics and}\\
\normalsize{Department of Physics, Fuzhou University, Fuzhou 350108, China}\\
\normalsize{$^{2}$Department of Physics, Taiyuan Normal University, Jinzhong 030619, China}\\
\normalsize{$^{3}$Beijing National Laboratory for Condensed Matter Physics}\\
\normalsize{Institute of Physics, Chinese Academy of Sciences, Beijing 100190, China}\\
\normalsize{$^{4}$College of Physics and Electronic Engineering}\\
\normalsize{Chongqing Normal University, Chongqing 401331, China}\\
\normalsize{$^{5}$School of Physics $\&$ Information Science,}\\
\normalsize{Shaanxi University of Science and Technology, Xi'an 710021, China}
\\
\normalsize{Correspondence: $^\ast$ eurake27hywang@fzu.edu.cn,$^\dagger$ wlqx@cqnu.edu.cn,$^\ddagger$ xfzhang@sust.edu.cn}
}
\date{}
\begin{document}



\baselineskip24pt


\maketitle

\section*{Abstract}
We investigate the localization behavior of two-particle Hubbard model in the presence of non-reciprocal tunneling and non-Hermitian bound states can be obtained with strong repulsive  interaction. Remarkably,  the interaction induced bound state localization (BSL) can compete with non-Hermitian skin effect (NHSE) and give rise to diverse density profiles. Via the quantum scattering methods in the center of mass frame, the system can be mapped to an effective two dimensional (2D) lattice with the two-particle interaction contributing to a defective line. For the bound states of the largest eigen-energy, in contrast to the Hermitian cases, where the maximal localization center is  pinned around the center of lattice, NHSE can lead to a faded diagonal line localization. For the unbound scattering states, unlike the single corner localization in the 2D Hatano-Nelson model, interaction can force the total localization to split into multiple centers.       To make the system topological nontrivial, we further include terms taking the form of two-photon tunneling and the non-Hermitian photon bound pairs are also observed, which demonstrates a competition among  NHSE, BSL and edge localization.   Finally, we propose the experimental simulations via the platforms of electrical circuits. Our works shed light on the crossover study of quantum optics and non-Hermitian many body physics.

\section*{INTRODUCTION}
The open quantum system with non-reciprocal tunneling or  onsite dissipation  features non-Hermiticity and can give birth to the presence of NHSE, where not only the topological edge modes but all the bulk states are piled up at the  end of the lattice \cite{Ueda,RevModPhys.93.015005,PhysRevLett.121.086803,PhysRevLett.125.126402,PhysRevLett.123.066404,PhysRevA.106.052216,PhysRevA.104.022215,PhysRevLett.133.076502,PhysRevLett.123.170401,PhysRevB.109.045122,ObserveNHSE1,PhysRevLett.124.086801,HFWANG,PhysRevResearch.6.023135,Nat2,PhysRevLett.132.096501,PhysRevB.102.205118,PhysRevLett.125.186802,PhysRevResearch.6.013213,PhysRevB.108.235422,PhysRevB.103.014203,PhysRevB.109.165127,PhysRevLett.128.157601,PhysRevB.105.245143,PhysRevB.100.054301,PhysRevLett.129.264301,PhysRevB.104.L241402}. Consequently,  the advent of non-Hermitian physics  greatly challenges the conventional understanding of optical and condensed matter systems \cite{PhysRevLett.120.146601,PhysRevLett.112.203901,PhysRevResearch.6.023004,nanophotonics,Stefano,PhysRevB.104.125416,PhysRevB.97.121401,PhysRevLett.121.073901,PhysRevA.100.062118,Nat4,PhysRevB.106.134112,PhysRevLett.127.140504,Communphys,Nat5,Natr2,PhysRevB.98.165435,Science2,PhysRevB.103.165123,PNAS,Science,PhysRevA.105.022215,PhysRevB.108.064311,PhysRevB.101.235150,Nature,PhysRevA.98.042114}.  For example, the existences of in-gap topological  boundary modes are usually  characterized by the nonzero bulk topological invariants, of which the phenomenon is dubbed as bulk-boundary correspondence. However, NHSE can break this rule and the degenerate edge states may get collapsed \cite{YeXiong,Natr,PhysRevLett.124.056802,ETI,PhysRevB.106.L041302,PhysRevLett.121.026808,PhysRevB.105.205402,PhysRevResearch.4.L012006,PhysRevResearch.6.L032067,PhysRevLett.129.180401,PhysRevB.109.155137,PhysRevB.108.184304,QST,PhysRevB.110.085431,PhysRevB.105.165137,PhysRevB.107.115146,PhysicaA,PhysRevResearch.5.L022046,PhysRevA.108.052210,PhysRevX.9.041015,PhysRevResearch.5.033058,PhysRevLett.126.010401,PhysRevB.99.245116,PhysRevA.109.062220,PhysRevB.103.075126,PhysRevLett.116.133903,PhysRevLett.129.113601,PhysRevLett.123.246801,PhysRevB.107.L201116}.  From the perspective of quantum transport, NHSE can also be explained by the imaginary fluxes pierced through the lattice and  be indicated via the non-zero area enclosed by the real and imaginary momentum space energy spectrum.

Quantum interaction has always been one of the most intriguing topics in the study of quantum information as well as modern condensed matter physics, which  plays a pivotal role in determining the quantum phases and the density profiles. In particular, for the two-particle Hubbard model, apart from the continuum state, bosons can be bound together to form a stable dimmer even in the conditions of strong repulsive on-site interaction, corresponding to which the localization can be termed as BSL   \cite{RevModPhys.80.1355,RevModPhys.84.299,NatureBSL,PhysRevB.40.506,PhysRevA.110.L020201,JPB,PhysRevA.106.043315,JPCM,PhysRevA.81.042102,PhysRevA.78.033611,PhysRevB.109.235412,PhysRevLett.133.140202}.  However, despite the achieved results so far, how the NHSE singly obtained with non-reciprocal quantum tunneling will compete with the interaction induced BSL remains rarely reported.

In this paper, we focus on the non-Hermitian two-particle Hubbard model with  repulsive on-site interaction. Numerically, through the quantum scattering methods in the center of mass frame, the system can be mapped to a 2D Hamiltonian with non-reciprocal tunneling in both directions. Meanwhile, the on-site interaction will result in a defective line, and around the vicinity of which, two-body bound states can be formed. For the sake of completeness, we first review BSL in Hermitian cases. In periodical boundary conditions (PBCs), bound state is equally distributed along the mapped defective line. While for open boundary conditions (OBCs), truncation effects suggest that Hermitian bound state of the largest eigen-energy gets the maximal localization pinned around the center of the lattice. As non-Hermiticity is turned on, NHSE can only survive in the conditions of OBCs and will cooperate with BSL, giving birth to exotic localization behaviors. Specifically, for the cases that NHSE is moderately coincided with the BSL, bound states of different energies appear a faded diagonal line localization with a single maximum.  When considering the unbound scattering states, 2D NHSE tends to form the corner localization with a single center. However, interaction can ruin this by splitting the total localization into multiple centers. Such phenomenon can be explained as a result of the competition among multiple scaling factors.

Since Hubbard model shares fruitful applications in quantum optics, we also make a further analysis by including terms taking the form of two-photon tunneling. In the strong interacting limit, the system is found to be topological nontrivial and  the effective model projected along the diagonal line mimics the non-Hermitian Su-Schrieffer-Heeger (NH-SSH) chain. Meanwhile, a cooperation among BSL, NHSE and edge localization is  demonstrated, where the topological nontrivial two-photon bound pair is observed at either the left-down or the right-up corner. Finally, we propose the experimental simulations through the platforms of electrical circuits.

\section*{RESULTS}

\subsection*{NHSE in the non-reciprocal two-particle Hubbard Model}

Typically, the 1D non-Hermitian Hubbard model can be described as the following
\begin{equation}
\hat{H}=\sum_j [-J_1 \hat{\alpha}^{\dagger}_j\hat{\alpha}_{j+1}-J_2 \hat{\alpha}^{\dagger}_{j+1}\hat{\alpha}_{j}]+\frac{U}{2} \sum_j\hat{n}_j (\hat{n}_j-1),
\end{equation}
where $\hat{\alpha}^{\dagger}_j (\hat{\alpha}_j)$ is the  creation (annihilation) operator on the $j$th site and $\hat{n}_j=\hat{\alpha}^{\dagger}_j\hat{\alpha}_j$.   $J_{1}\neq J_{2}$ measures the non-reciprocal tunneling and $U \in R, U>0$ denotes the strength of on-site repulsive interaction. As we focus on two distinguished particles $a,b$ trapped in the lattice systems above, the first quantized Hamiltonian can be recast as following
\begin{eqnarray}
\begin{aligned}
H=&\sum_j(-J^{1}_a|a_j\rangle\langle a_{j+1}|-J^2_a|a_{j+1}\rangle\langle a_{j}|-J^{1}_b |b_j\rangle\langle b_{j+1}|-J^2_b|b_{j+1}\rangle\langle b_{j}|)\\
&+U\sum_m |a_m,b_m\rangle\langle a_m,b_m|,
\end{aligned}
\end{eqnarray}
where $|a(b)_j\rangle$ denotes the state with a single $a(b)$ particle at the $j$th site of lattice. The schematic picture of the system is shown in Figure 1a. To analyze  localization behaviors,  the two-particle state vector can be expanded as $|\Psi\rangle=\sum_{w,v}\psi(a_w,b_v)|a_w,b_v\rangle$, which describes the opportunities for locating  $a$ particle at $w$th site and  $b$ particle at $v$th site. Through the Schrodinger equation, the following recurrence relation can be obtained
\begin{eqnarray}
\begin{aligned}
E\psi(a_w,b_v)=&-J^{1}_b \psi(a_w,b_{v+1})-J^{2}_b \psi(a_w,b_{v-1})-J^{1}_a \psi(a_{w+1},b_{v})\\
&-J^2_a\psi(a_{w-1},b_v)+U \psi(a_w,b_v)\delta_{w,v}.
\end{aligned}
\end{eqnarray}
For the chain of length $L$, Eq.(3) suggests a mapping from the 1D two-particle Hamiltonian to a 2D single body effective form, and during the process of which, the size of Hilbert space, $L^2$, remains unchanged. Such mapping formalism is also frequently applied in refs [86-91].  Meanwhile, specific to our model, it should also be pointed out that the collisional interaction $U$ shares the same effect of a defective line and  $J^{1,2}_{a,b}$ depicts the non-reciprocal tunnelings in $x$ and $y$ directions.
\begin{figure*}[t]
\centering
\includegraphics[width=0.97\textwidth,height=0.452\textheight]{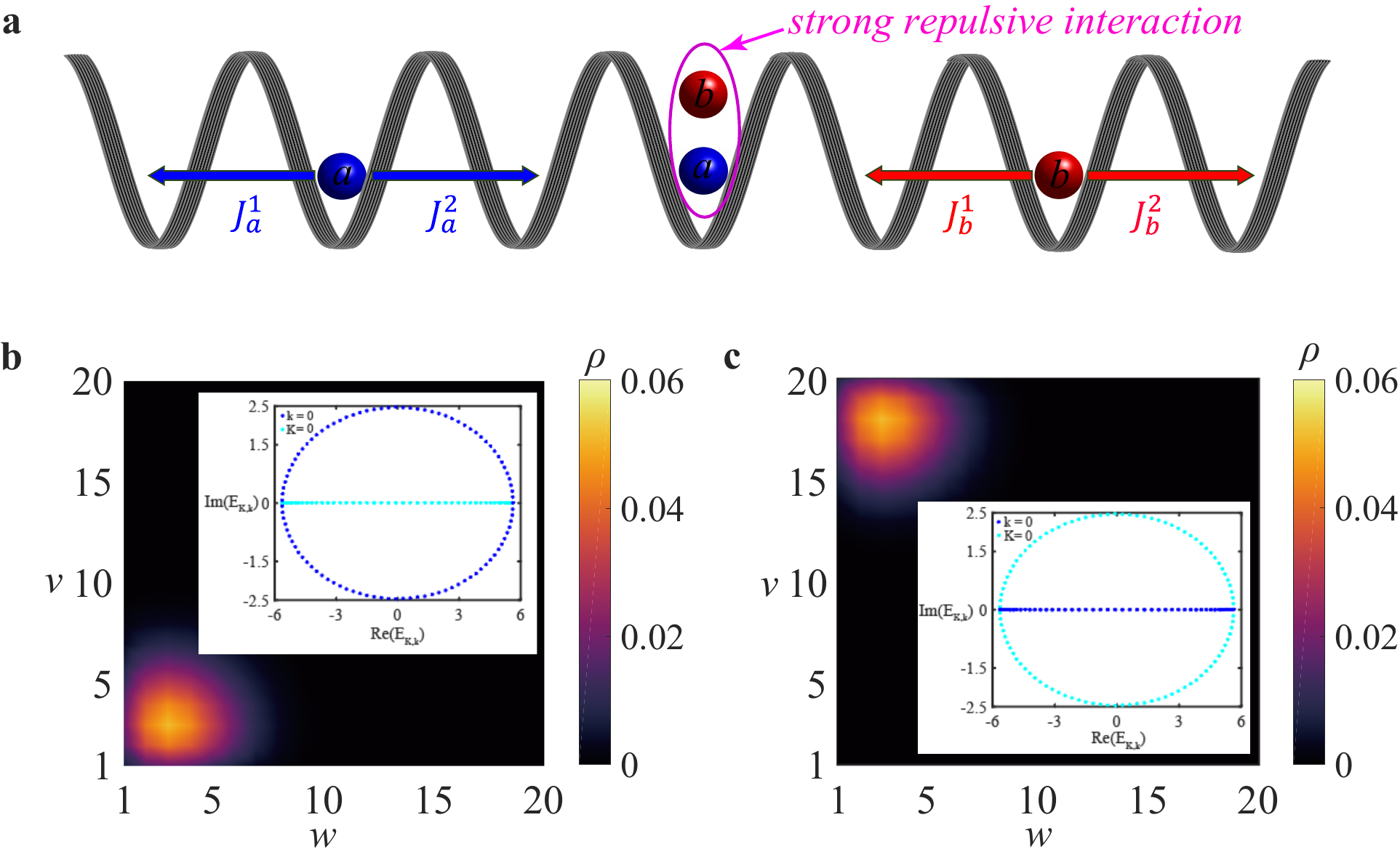}
\caption{\textbf{a} The schematic picture of the non-reciprocal two-particle Hubbard model. Atoms can be bound into dimmers by the strong repulsive interaction.  \textbf{b} The localization strength is measured via $\rho=|\psi(a_w,b_v)|^2$. When $J^0_a=J^0_b$, $\beta_1=\beta_2>0$, eigen-states tend to be piled up at the left-down corner.  Momentum space energy spectrum $\mathrm{Re(E_{K\neq 0,k=0})}$ and $\mathrm{Im(E_{K\neq 0,k=0})}$ encloses non-zero area. \textbf{c} For $J^0_a=J^0_b$ and $\beta_1=-\beta_2<0$, NHSE is manifested by the left-up corner localization. Non-zero winding in energy spectrum is observed by setting $K=0,k\neq0$. } \label{fig1}
\end{figure*}

The exact solutions of two-body energy spectrum can be obtained via the quantum scattering methods in the center of mass frame. Here, we rewrite $\psi(a_w,b_v)=\sum_{K} e^{iKR}\psi_K(r)$, where $R=\frac{w+v}{2}$ features the mass center and $r=w-v$ manifests the relative position. $K$ represents the Bloch wave vector of the dimmer bound pair. Then Eq.(3) can be redescribed as following
\begin{eqnarray}
\begin{aligned}
E_K \psi_K(r_j)&=\!-\!J^{1}_b e^{iK\frac{d}{2}}\psi_K(r_{j-1})\!-\!J^{2}_b e^{-iK\frac{d}{2}}\psi_{K}(r_{j+1})\!-\!J^{1}_a e^{iK\frac{d}{2}}\psi_K(r_{j+1})\\
&-J^2_a e^{-iK\frac{d}{2}}\psi_K(r_{j-1})+U\psi_K(r_j)\delta_{r_j,0}.
\end{aligned}
\end{eqnarray}
Considering the non-Hermiticity, the following declaration can be made
\begin{eqnarray}
\begin{aligned}
J^{0}_{b(a)}=\sqrt{J^1_{b(a)} J^{2}_{b(a)}},\quad  e^{\beta_{1(2)}}=\sqrt{\frac{J^1_{b(a)}}{J^2_{b(a)}}}, \quad \beta_1=\frac{d}{2}\gamma_1, \quad \beta_2=\frac{d}{2}\gamma_2.
\end{aligned}
\end{eqnarray}
Through Eq.(5), the generalized Bloch wave vector in the complex plane can be defined as $\theta_{1(2)}=K-i\gamma_{1(2)}$ in correspondence. Hence, it can be concluded that the non-reciprocal tunneling $|J^{1}_{a(b)}|\neq |J^{2}_{a(b)}|$ will lead to $|e^{i\theta_{1(2)}}|\neq 1$ inducing the NHSE of the two-body bound  states. Besides, Eq.(4) can also be redepicted as
\begin{eqnarray}
\begin{aligned}
J_L \psi_K(r\!_{j\!-\!1})\!\!+\!\!J_R\psi_K(r\!_{j\!+\!1})\!\!+\!\!U\psi_K(r\!_j)\delta_{r\!_j,0}\!\!=\!\!E_K \psi_K(r\!_j)\!,
\end{aligned}
\end{eqnarray}
where $J_L=-J^{0}_{b} e^{i\theta_1\frac{d}{2}}-J^{0}_a e^{-i\theta_2\frac{d}{2}}, J_R=-J^{0}_b e^{-i \theta_1\frac{d}{2}}-J^0_a e^{i\theta_2\frac{d}{2}}.$ In general, $J_L\neq J_R $ unless $\frac{\sin{(\theta_1 \frac{d}{2})}}{\sin{(\theta_2 \frac{d}{2})}}=\frac{J^0_a}{J^0_b}$.

\subsection*{Localization in the non-interacting  limit}

The non-interacting limit $U\rightarrow 0$  is equivalent to get $a,b$ particles decoupled, and the energy spectrum appears as $E^{U_0}_{K,k}=E[q_a]+E[q_b]$, where $E[q_{a(b)}]=-J^0_{a(b)}[e^{iq_{a(b)}d+\beta_{2(1)}}+e^{-iq_ {a(b)}d-\beta_{2(1)}}]$ is the non-Hermitian single-body momentum space spectrum when only $a(b)$ particle exists. By denoting $K=q_a+q_b, k=\frac{1}{2}(q_a-q_b)$, the exact form of  wave function can be given as $\psi_K(r_j)=e^{ikr_j}$, and eigen-energy in Eq.(6) matches the $E^{U_0}_{K,k}$ above. At this stage, the NHSE can be identified by the windings of $E^{U_0}_{K,k}$, which is
\begin{equation}
w=\int^{\pi}_{-\pi} \frac{dk(K)}{2\pi}\partial_{k(K)} \mathrm{arg} (E^{U_0}_{K,k}).
\end{equation}
To illustrate, we consider two concrete examples. For the case $\beta_1=\beta_2, J^0_a=J^0_b$, although $J_L=J_R$, non-zero area enclosed by $E^{U_0}_{K,k}$  with fixed $k=0$ is observed [see Figure 1b], where $w=-1$ and all states are localized at the right-up or the left-down corner depending on the sign of $\beta_{1(2)}$.  Considering the case $\beta_1=-\beta_2, J^{0}_a=J^{0}_b$, the NHSE is preserved, which can be indicated by $w=1$ with fixed $K=0$ [see Figure 1c], and all states are pinned around the left-up or the right-down corner.

\subsection*{Localization with the finite interaction strength}
\begin{figure}[tb]
\centering
\includegraphics[width=0.90\textwidth,height=0.27\textheight]{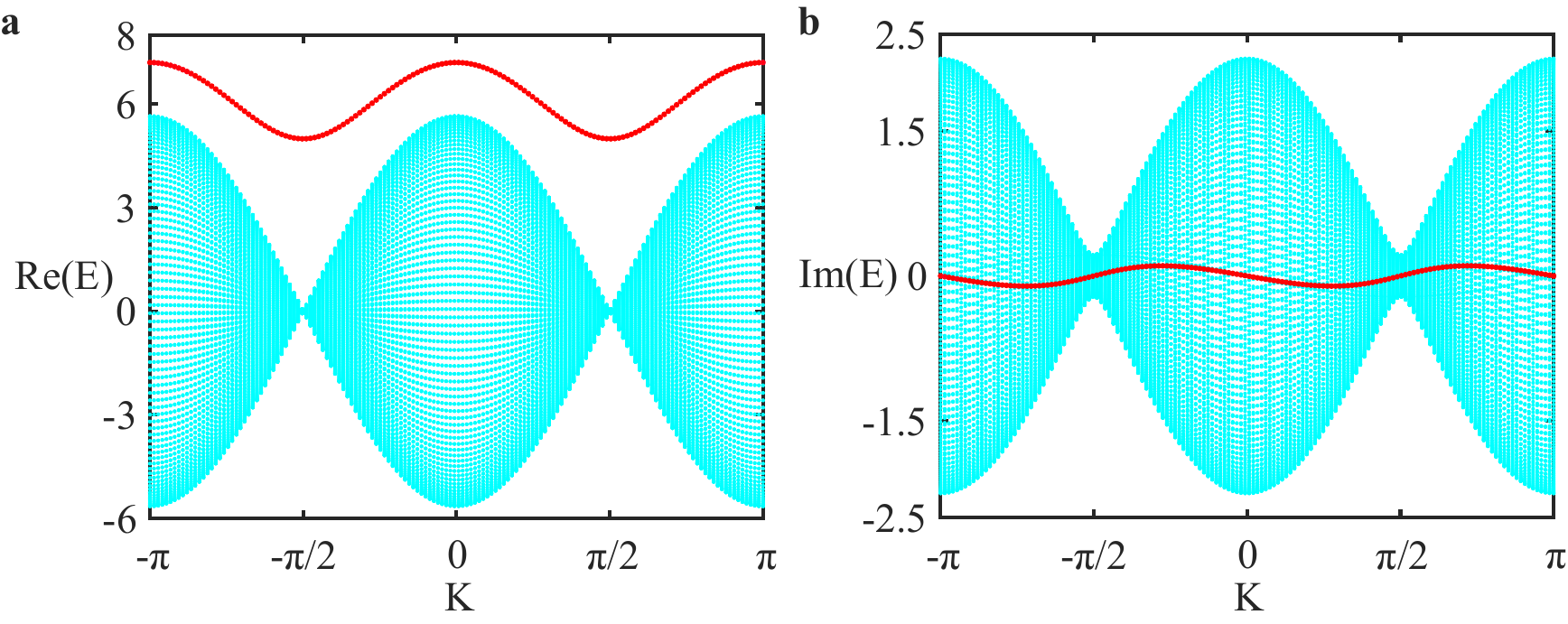}
\caption{ Energy spectrum as a function of the center of mass momentum $K$ at finite interaction strength $U=5$ and $ J^1_a=1,J^2_a=J^1_b=1.6, J^2_b=1.1$. Red lines correspond to the bound states and cyan lines depict the unbound scattering states.  } \label{fig2}
\end{figure}

As the interaction strength is enlarged towards infinity, the bound states localized at $r_j=0$, where the two particles $a,b$ are pinned on the same site,  will inevitably appear. Meanwhile, the energy spectrum  shrinks towards the flat band $E^{U}_{K,k}=U$. Obviously, according to  the discussion above, it can be expected that enlarging the two-body interacting strength from zero to a finite value will drastically change the localization behavior. To get an analytical result, we consider the lattice Green function operator $G$, which should be defined via the following equations
\begin{equation}
 \langle \psi_K(r')|(H_0-E)G| \psi_K(r)\rangle=\delta_{r,r'}.
\end{equation}
Here, $H_0$ describes the tunneling part in Eq.(6) and $G$ can be expanded as
\begin{equation}
G=\sum_{r,r'} g(K; r,r')|\psi_K(r)\rangle\langle \psi_K(r')|,
\end{equation}
where $g(K;r,r')$ is termed as the lattice Green function and  is the  solution to
\begin{equation}
\delta_{r,r'}=-E g(K,r,r')-J_L g(K,r-1,r')- J_R g(K,r+1,r').
\end{equation}
To proceed, the eigen-states for the  finite interaction strength can be derived  via the  Lippmann-Schwinger equation
\begin{equation}
|\psi_K(r)\rangle=|\psi^{U=0}_K(r)\rangle+GU|\psi^{U}_K(r)\rangle.
\end{equation}
Correspondingly, the non-Hermitian bound state spectrum is shown with red lines in Figures 2a-2b.

\begin{figure}[t]
\centering
\includegraphics[width=0.90\textwidth,height=0.27\textheight]{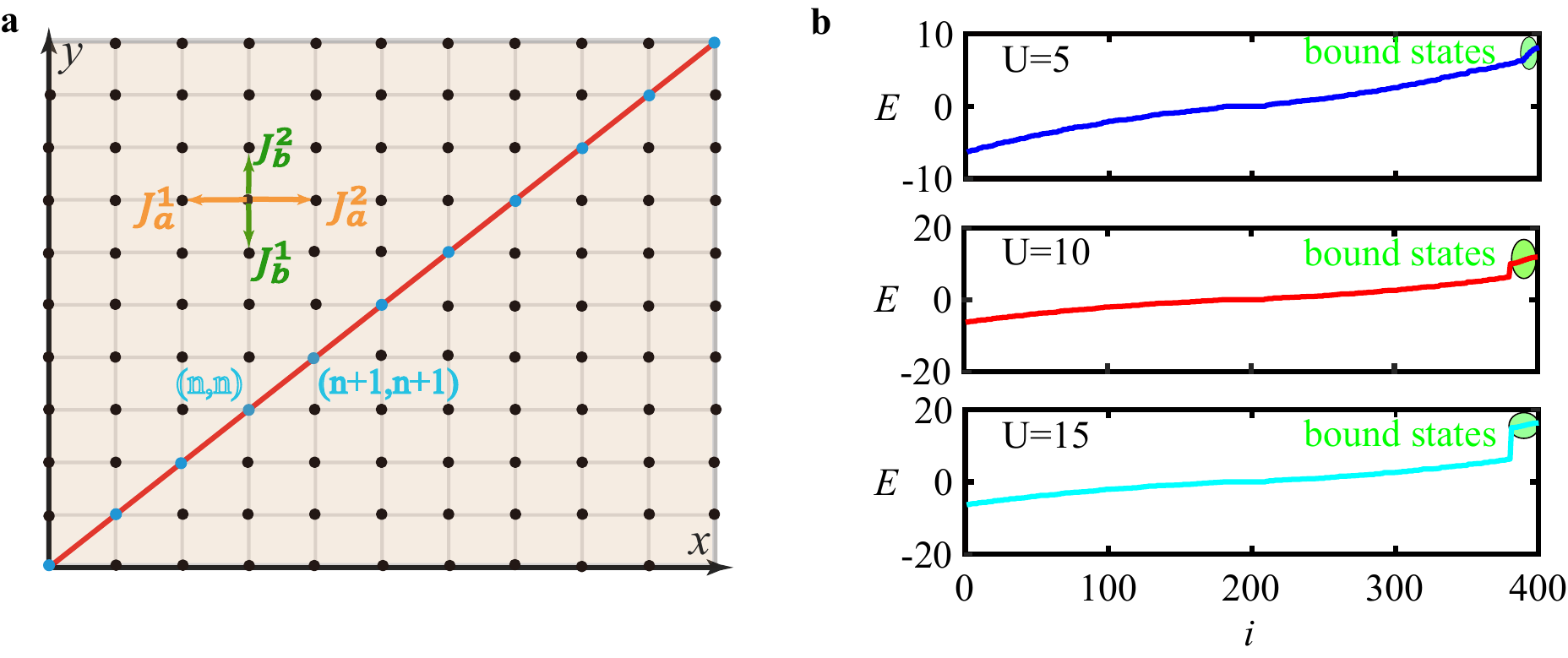}
\caption{\textbf{a} Eq.(3) can be viewed as the iteration function of a effective 2D single-body model. $J_{a,b}$ correspond to tunnelings in the horizontal and vertical directions. Two-body interaction $U$ can be reinterpreted as a defective line. \textbf{b} For PBCs and $J^1_a=J^2_a=J^1_b=J^2_b=1.6$, the bound states shaded green region with $U=5,10,15$ differ from the unbound scattering states with a jump in eigen-energy.} \label{fig2}
\end{figure}
The localization behavior of the bound state is more explicitly shown in the position space, which can be numerically obtained by a direct diagonalization of the 2D effective Hamiltonian schematically shown by Figure 3a.   Specifically, we begin by considering the bound states in PBCs, which can be fulfilled by adding the following tunneling to the Hamiltonian in Eq.(2)
\begin{equation}
H_{\textrm{PBCs}}=-J^1_a |a_L\rangle\langle a_1|-J^2_a |a_1\rangle\langle a_L|-J^1_b |b_L\rangle\langle b_1|-J^2_b |b_1\rangle\langle b_L|,
\end{equation}
and before proceeding we review the prototypical localization  in Hermitian cases. Generally, the two-body bound states discussed here differ from the continuum states by a jump in eigen-energy, which are manifested by the green shaded region in Figure 3b. It can be seen that enlarging the interaction strength $U$ can get the bound states more distinguished. In Figure 4a, the localization of the bound state of the largest eigen-energy $E=11.88$ with interaction strength $U=10$ is demonstrated, where a uniform distribution along the diagonal line is illustrated by $\rho=|\psi|^2$. Such phenomenon means that the dimer bound pair can be found at any site of the original 1D lattice with equal possibility.  For the OBCs, truncation effect has to be considered and  the localization is more easily to be affected by the quantum tunneling. For example, in Figure 4b, we still focus on the bound state of the largest eigen-energy. Although the main distribution still lies at the diagonal line, it is no longer uniform and the dimmer bound pair is maximally localized around  the center of the lattice.

\begin{figure}[t]
\centering
\includegraphics[width=0.979\textwidth,height=0.39\textheight]{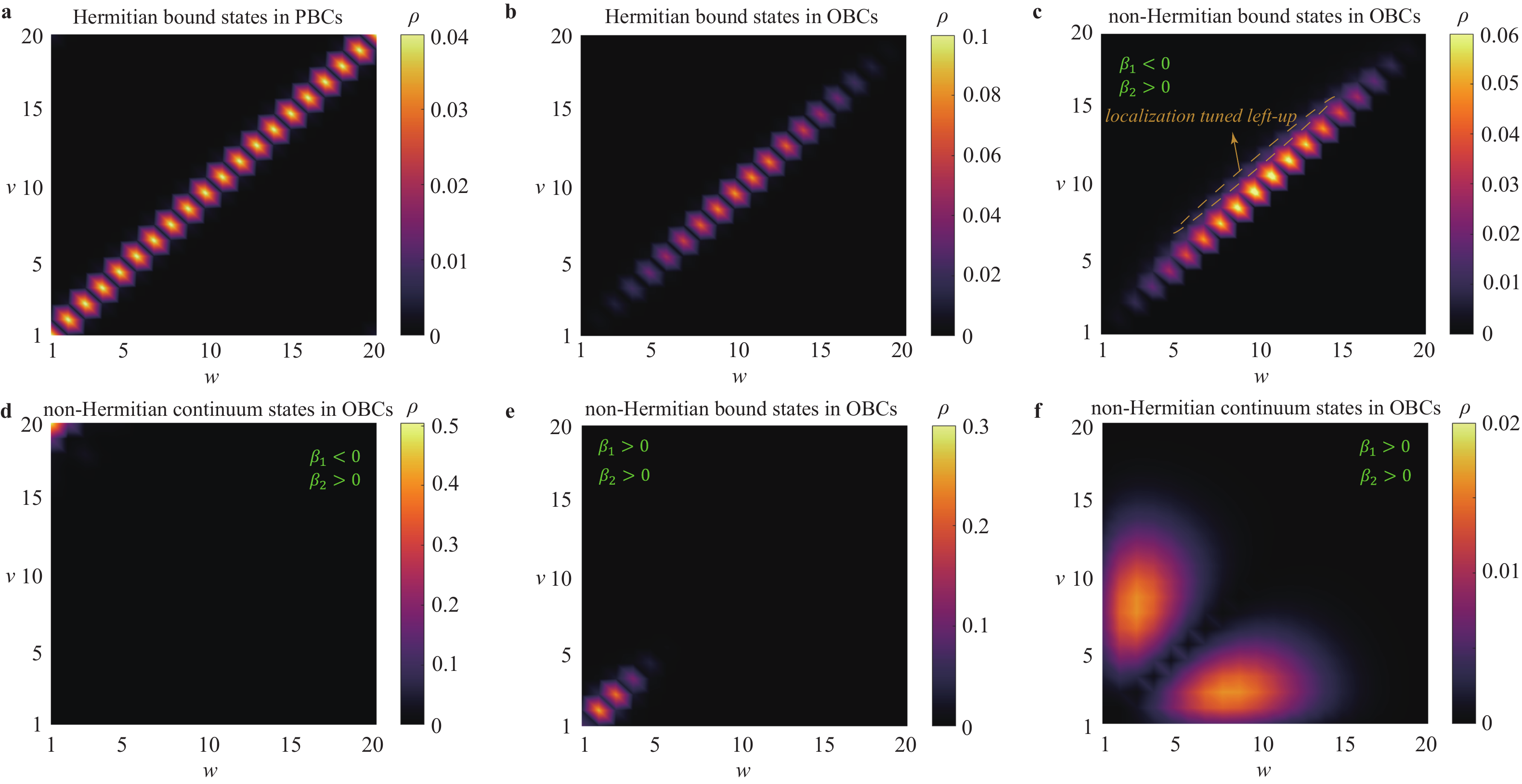}
\caption{\textbf{a} In Hermitian cases, setting $U=10,J^1_a=J^2_a=J^1_b=J^2_b=1.6$ and PBCs, bound state of the largest eigen-energy is uniformly distributed along the diagonal line. \textbf{b} For $U=10$ and OBCs, the bound state above is maximally localized around the center of the lattice.  \textbf{c} In non-Hermitian cases and $U=10$, $J^1_a=J^2_b=3.0, J^2_a=J^1_b=1.0$ indicates $\beta_1<0, \beta_2> 0$, localization of the bound state of the largest eigen-energy is partially tuned to the left-up side.  \textbf{d} Corresponding to the same parameters in \textbf{c}, the unbound scattering state is generally localized at the single left-up corner as predicted by the 2D NHSE. \textbf{e} Setting  $J^{1}_a=J^{1}_b=1.0, J^2_a=J^2_b=1.6, U=5$, bound state with eigen-energy $|E|=6.9$ shows a faded diagonal line localization with a single maximum. \textbf{f} Corresponding to the same parameters in \textbf{e}, non-Hermitian unbound scattering state of eigen-energy $|E|=4.9$ does not center around the single corner but the total localization splits into two parts. } \label{fig2}
\end{figure}

In non-Hermitian cases with the non-reciprocal tunneling, the localization can be greatly changed by the presence of NHSE. To begin with, we apply OBCs and set $J^0_a=J^0_b,\beta_1=-\beta_2$. In this scenario, 2D NHSE tends to predict the left-up (or the right-down) corner localization, which does not overlap with the BSL. If the 2D NHSE totally conquered the BSL, the two-particle wave function would satisfy $\psi(a_w,b_v),w=1(L),v=L(1)$ and no bound states ($w=v$) could be observed.  While the numerical results prove that the bound state can still survive with the total localization partially tuned towards the corner suggested by the NHSE. In Figure 4c, we set $J^1_a=J^2_b=3.0, J^2_a=J^1_b=1.0, U=10$, corresponding to which we have $\beta_1<0, \beta_2> 0$, and the localization of bound state of eigen-energy $E=12.15$ is demonstrated. Besides, the  unbound scattering states are maximally localized at the left-up corner as predicted by the 2D NHSE [see Figure 4d]. Considering states of different energies, transitions from bound states to the unbound scattering states can be quite sharp for the system of large enough size.

On the other hand, for $J^0_a=J^0_b,\beta_1=\beta_2$,  the 2D NHSE tends to predict the corner localization pinned at the right-up (or the left-down) side, which can moderately be coalesced with the BSL. In such circumstances, the localizations of bound and unbound states are greatly altered. Specifically, in Figure 4e, the bound state of eigen-energy $|E|=6.9$ cooperates with the non-reciprocal tunneling, giving rise to a faded diagonal line localization with a single maximum. Remarkably, unlike the previous cases, the unbound scattering states are more easily affected by the two-body interaction. For example, continuum state of the smallest eigen-energy $|E|=4.9$ does not localize at the single left-down corner, but splits into two fading parts [see Figure 4f].

\begin{figure}[t]
\centering
\includegraphics[width=0.88\textwidth,height=0.489\textheight]{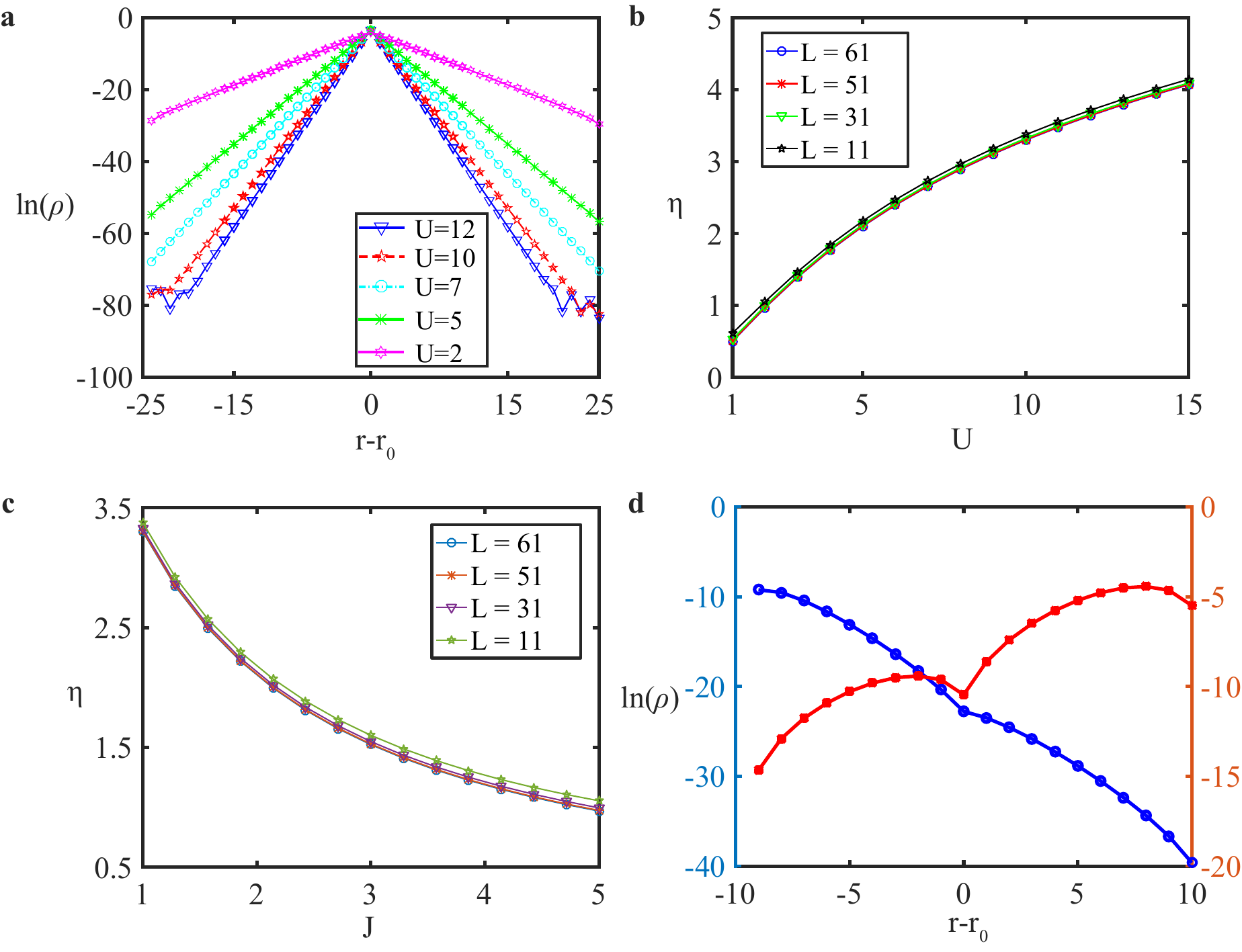}
\caption{\textbf{a} In Hermitian cases and $J^1_a=J^1_b=J^2_a=J^2_b=J=1.0$, the density profile $\ln(\rho)$ varies as a function of $|r-r_0|$ with fixed $v=\frac{L+1}{2}$, of which the slope manifests the scaling factor $\eta$. \textbf{b} With fixed $J=1$, scaling factor $\eta$ changes as a function of the two-body interaction strength $U$. \textbf{c} With fixed $U=10$, $\eta$ decays with the variation of tunneling amplitude $J$. \textbf{d} In non-Hermitian cases and $U=5$, different scaling factors shall be combined together in determining the localization behavior.  Blue line corresponds to parameters $J^1_a=J^1_b=5.0, J^2_a=J^2_b=1.0$. Red line is obtained with $J^1_a=J^1_b=1.0, J^2_a=J^2_b=1.6$. } \label{fig5}
\end{figure}

Results above can be explained by the competition between BSL and 2D NHSE. In detail, the BSL caused by the two-body interaction in Hermitian cases can be depicted as
$\rho^H=|\psi^{H}(a_w,b_v)|^2 \propto \mathcal{M}(r-r_0) \mathcal{N}(R-r_c)$,
where $r-r_0=0$ for $w=v$, and $R-r_c$ measures the distance away from the lattice center. Typically, $\mathcal{M}(r-r_0)=e^{-\eta|r-r_0|}$. $\eta$ is the localization scaling factor and is much larger than that in $\mathcal{N}(R-r_c)$, playing the leading role.  Meanwhile, $\eta$ varies as a function of $U$ and tunneling amplitude $J^0_a=J^0_b=J$ [see Figures 5a-5c], exhibiting no dependence on the size of system. Effects of $\mathcal{N}(R-R_c)$ is demonstrated in the methods part.   To proceed, it should also be noticed that the Hamiltonian in Eq.(1) can be made Hermitian by applying the following transformation on the basis
\begin{equation}
|a_w,b_v\rangle = e^{-\beta_1 v} e^{-\beta_2 w} |a_w,b_v\rangle'.
\end{equation}
The mapped Hermitian systems $H'=M^{-1} H M$ shares the same spectrum as the non-Hermitian one. The density profile of the non-Hermitian states can be extracted from its Hermitian counterpart and appears as following
\begin{equation}
\rho=|\psi(a_w,b_v)|^2\propto
\begin{cases}
e^{-(\eta+\beta_1)(r-r_0)} e^{(\beta_1+\beta_2)w}\quad \text{for $r>r_0$}\\
e^{-(\eta-\beta_1)(r_0-r)} e^{(\beta_1+\beta_2)w}\quad \text{for $r<r_0$}.
\end{cases}
\end{equation}

For the special case $\beta_1=\beta_2=\beta$, the scaling factor $\eta$ is changed to two exponents $e_1=\eta+\beta, e_2=\eta-\beta$ with the consideration of NHSE. For the conditions  $\eta\gg |\beta|$, Eq.(14) suggests the vanished intensity at $r\gg r_0$ and $e^{(\beta_1+\beta_2)w}$ will account for how the total localization is tilted towards the side depicted by the NHSE. Such  results are consistent with Figures 4d-4f. For the conditions $\eta\ll |\beta|$, NHSE plays the dominant role. In other circumstances,  the consequences of the interplay of the multiple scaling factors are manifested in Figure 5d, where the  maximum of $\ln(\rho)$ is tuned away from $r=r_0$, accounting  for the multiple (or two) localization centers in our two-body system.
\subsection*{Topological photon bound pairs in the extended Hubbard model}
In the study above, the system is conceived to be topological trivial and shares  potential applications in studying the distribution of photon bound pairs in the non-reciprocal waveguide arrays. To gain non-trivial topological properties, the Hubbard model can be expanded to be bipartite or to include terms taking the form of two-photon tunneling
\begin{equation}
H_{2P}= \sum_{m}P(a^{\dagger}_{2m+1}a^{\dagger}_{2m+1}a_{2m}a_{am}+h.c.),
\end{equation}
where $a_m$ denotes annihilating a photon at the $m$th site.  Now, we fix $J^1_a=J^1_b=J_1,J^2_a=J^2_b=J_2$ and by reassuming two-particle wave function to be $\psi=\sum_{m,n} \eta_{m,n} |a_m,a_n\rangle$, the bulk iteration function in Eq.(3) is changed to the form
\begin{eqnarray}
\begin{aligned}
(E-U)\eta_{2m,2m}=&-J_1\eta_{2m-1,2m}-J_2\eta_{2m+1,2m}+P\eta_{2m+1,2m+1}.
\end{aligned}
\end{eqnarray}
\begin{eqnarray}
\begin{aligned}
(E-U)\eta_{2m+1,2m+1}\!=&\! -J_1\eta_{2m,2m+1}\!-\!J_2\eta_{2m+2,2m+1}+P\eta_{2m,2m}.
\end{aligned}
\end{eqnarray}
\begin{eqnarray}
\begin{aligned}
E\eta_{m,n}=&-J_1(\eta_{m,n+1}+\eta_{m+1,n})-J_2(\eta_{m,n-1}+\eta_{m-1,n}).
\end{aligned}
\end{eqnarray}

\begin{figure}[t]
\centering
\includegraphics[width=0.90\textwidth,height=0.27\textheight]{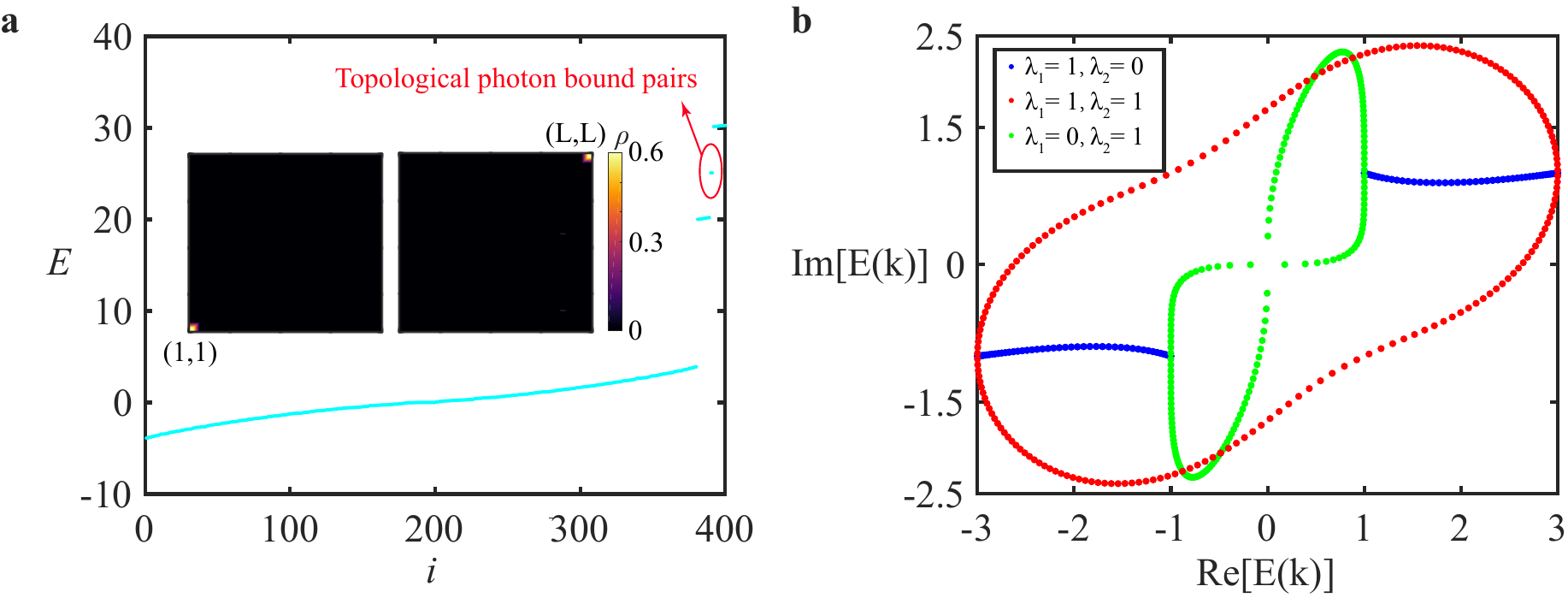}
\caption{\textbf{a} For the extended two-particle Hubbard in Hermitian conditions $J_1=J_2=1, U=25, P=5,L=20$, topological nontrivial photon bound pairs are circled out with red lines, which are pinned at the left and right ends of the diagonal line. \textbf{b} Characterization of the NHSE by the windings of energy spectrum. For $P=1+i$, non-zero windings are observed with $\lambda_1=\lambda_2=1$. The cases $\lambda_2=0$ corresponds to $\beta=1$ and $\lambda_1=0$ indicates $\beta=-1$, both being absence of skin effects. } \label{fig4}
\end{figure}

In the strong interacting limit $U\gg J_1,J_2$, following the discussion above, $\eta_{m,m}$ plays the dominant role, and other components $\eta_{m,p}$, where $|p-m|\gg1$, can be treated as perturbations. Thus, Eqs.(16) and (17) will result in
\begin{eqnarray}
\begin{aligned}
&\left(E-U-\frac{J_T}{2}\right)\eta_{2m,2m}=\left(P+\frac{J_2}{2}\right)\eta_{2m+1,2m+1}+\frac{J_1}{2}\eta_{2m-1,2m-1}
\end{aligned}
\end{eqnarray}
\begin{eqnarray}
\begin{aligned}
\left(E-U-\frac{J_T}{2}\right)\eta_{2m+1,2m+1}=\left(P+\frac{J_1}{2}\right)\eta_{2m,2m}+\frac{J_2}{2}\eta_{2m+2,2m+2}
\end{aligned}
\end{eqnarray}
where $J_T=J_1+J_2$. It can be seen that Eqs.(19) and (20) simulate the bulk states of  non-Hermitian SSH model, and $\frac{J_{2(1)}}{2}$ mimics the intra-cell leftwards (rightwards) tunneling. $(P+\frac{J_{2(1)}}{2})$ is regarded as the inter-cell leftwards (rightwards) tunneling. In the momentum space, the effective model is depicted as
\begin{eqnarray}
\begin{aligned}
H_{eff}=&\left(U+\frac{J_T}{2}\right)I+[\lambda_1+(\lambda_1+P)\cos{k}-\!i\lambda_2\sin{k}]\sigma_x +[-i\lambda_2+(P+\lambda_1)\sin{k}\\
&+i\lambda_2\cos{k}]\sigma_y,
\end{aligned}
\end{eqnarray}
where $\lambda_1=\frac{J_1+J_2}{4},\lambda_2=\frac{J_1-J_2}{4}$. Topological phase transitions can be decided by the gap closing points, which take place at $|P|=|2\lambda_1|$, and different topological phases are characterized  by the winding number
\begin{equation}
W=\frac{1}{2\pi}\int^{2\pi}_0 [d_y(k)\partial_k d_x(k)-d_x(k)\partial_k d_y(k)] dk,
\end{equation}
where $d_{x,y}(k)$ are coefficients before Pauli matrices $\sigma_{x,y}$ respectively. In Figure 6a, when $\lambda_2=0$, the system retains Hermitian and topological nontrivial two-photon pairs lie within the gaps of bound state spectrum, which are localized  at the left-down and the right-up corners. Indeed, considering the two-dimensional effective model, Eqs.(19-20) suggest that two-photon tunneling contributes to the staggered coupling along the diagonal line, sharing  the same effect of  the $\pi$ flux in SSH model. As we turn on the non-Hermiticity, two corner modes are force towards a single side.  In Figure 6b, the  skin effects of the two-photon bound states are identified  by the windings of $\{\mathrm{Re}[E(k)],\mathrm{Im}[E(k)]\}$, and it is presented that the complex two-photon tunnelings are not necessary in constructing the skin effects, but will only change the shape of the circling energy spectrum.
\subsection*{Experimental simulation of the non-reciprocal Hubbard model}
To realize the non-reciprocal two-particle Hubbard model, we propose to utilize the electrical circuit lattices, of which the methods have frequently been applied in simulating diverse condensed matter systems \cite{Nat3,NatCommun,Research,PhysRevApplied.20.064042,PhysRevB.109.115406,PhysRevB.109.115407,PhysRevB.100.201406,PhysRevB.107.245114,PhysRevLett.122.247702,PhysRevResearch.2.033052,PhysRevB.109.L241103,PhysRevB.107.085426,PhysRevResearch.3.023056,PhysRevB.99.201411}. Before proceeding, we shall reemphasize that the applicability of this approach is based on the mathematically rigorous mapping of the two-body  problem to a classical setup of higher dimensionality. Specific to our model, the circuit simulation is illustrated in Figure 7, which elaborates the iterative relation of the Hamiltonian in Eq.(2). In detail, the current flowing through the $(m,n)$ node of circuit lattice is governed by the Kirchhoff's law
\begin{equation}
I_{(m,n)}=\sum_p L_{(m,n),p} V_{(m,n)},
\end{equation}
where $V_{(m,n)}$ denotes amplitude of the voltage potential. $L_{(m,n),p}$ describes the net impedances of all the electrical elements linked to the $(m,n)$ node, which is also termed as  circuit Laplacian. Considering our systems, the current on the $(m,n),m\neq n$ node is characterized by
\begin{eqnarray}
\begin{split}
i\omega C_0\! V_{m,n}&\!=\!R_1(V_{m+1,n}\!\!-\!\!V_{m,n})\!\!+\!\!R_2(V_{m-1,n}-V_{m,n})\\
&\!+\!R_3(V_{m,n+1}\!-\!V_{m,n})\!+\!R_4(V_{m,n-1}\!-\!V_{m,n}),
\end{split}
\end{eqnarray}
where it is set
\begin{eqnarray}
\begin{aligned}
& R_1=\frac{R_a-iR^2_a\omega L_2}{1+R^2_a \omega^2 L^2_2}, \quad R_2=\frac{-R_a-iR^2_a\omega L_2}{1+R^2_a \omega^2 L^2_2},\\
& R_3=\frac{R_b-iR^2_b\omega L_1}{1+R^2_b \omega^2 L^2_1}, \quad R_4=\frac{-R_b-iR^2_b\omega L_1}{1+R^2_b \omega^2 L^2_1}.
\end{aligned}
\end{eqnarray}
\begin{figure}[t]
\centering
\includegraphics[width=0.72\textwidth,height=0.38\textheight]{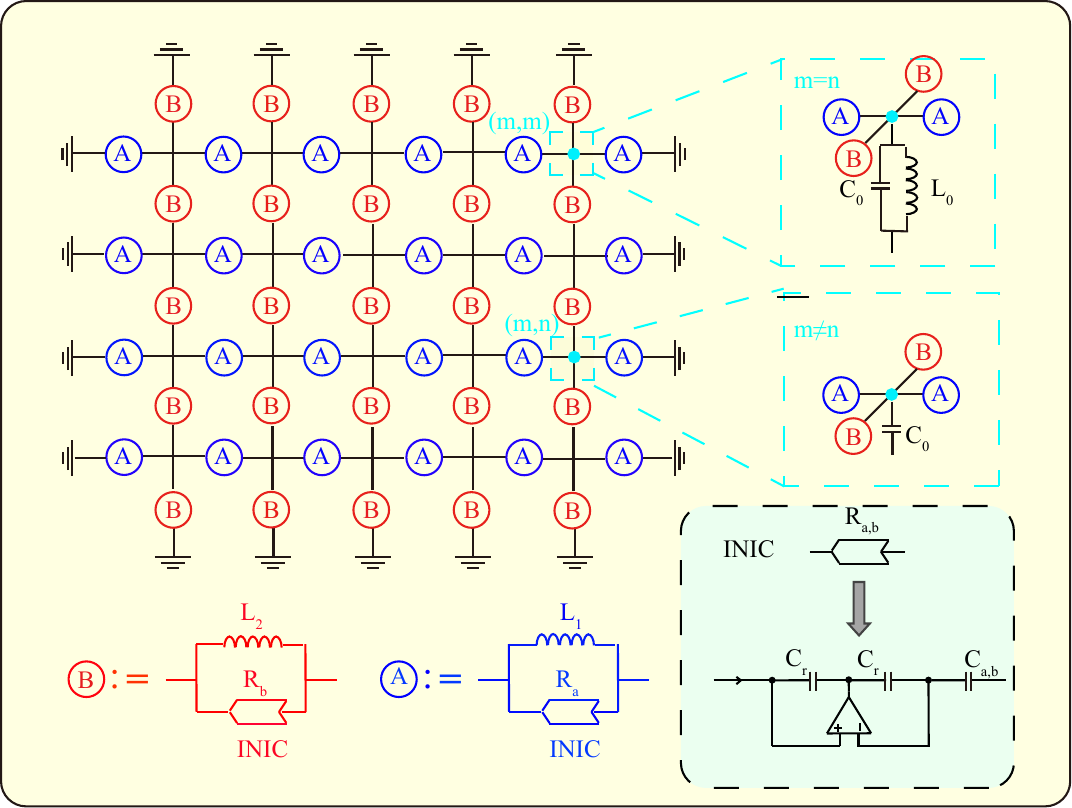}
\caption{ The schematic picture for the experimental realization of the non-reciprocal Hubbard model. The cyan box depicts the side view of the circuit lattice and the diagonal line nodes differ from others by the presence of extra inductors $L_0$. } \label{fig5}
\end{figure}
By keeping $R_{1,2}=i\omega C_0 J^{1,2}_{a}, R_{3,4}=i\omega C_0 J^{1,2}_b$, the non-reciprocal tunneling parts in Eq.(2) can be simulated. Here, $R_{a,b}$ measure the absolute amplitude of the negative impedance converter with current inversion (INIC), which is the central part for inducing non-Hermiticity. In detail, the impedance of INIC depends on the direction of the current. For the rightwards current, $R_{a,b}$ take positive values. While for the leftwards current, the impedances reverse to be negative. The detailed structures of INIC are illustrated in ref [102] and a modified version is shown by the green shaded region in Figure 7, where $R_{a,b}=i\omega C_{a,b}$.   To synthesize the two-body interacting part, the current on the diagonal node $(m,m)$ is given by
\begin{eqnarray}
\begin{aligned}
i\omega C_0 V_{m,m}&\!=\! R_1(V_{m+1,m}\!-\!V_{m,m})\!+\!R_2(V_{m-1,m}\!-\!V_{m,m})\\
&\!+\!R_3(v_{m,m+1}-V_{m,m})\!+\!R_4(V_{m,m-1}-V_{m,m})-\frac{1}{i\omega L_0} V_{m,m}.
\end{aligned}
\end{eqnarray}
By comparing Eq.(26) with Eq.(3), it can be derived that the two-body interaction strength is elaborated through $U=\frac{1}{\omega^2 L_0 C_0}$.
\section*{DISCUSSION}

We have explored the exotic localization behavior of the non-reciprocal two-particle Hubbard model via the methods of mapping the original Hamiltonian into a two-dimensional effective model. For PBCs, the bound states are uniformly distributed along the diagonal line. While for OBCs, the BSL caused by the interactions can coexist and compete with NHSE. For the cases that BSL and NHSE lying on the different sides, localization of the bound state is partially tuned towards direction suggested by the NHSE. For the cases that BSL and NHSE are moderately coalesced, the total localization appears to be a faded diagonal line with a single maximum.

\begin{figure*}[tb]
\centering
\includegraphics[width=0.97\textwidth,height=0.179\textheight]{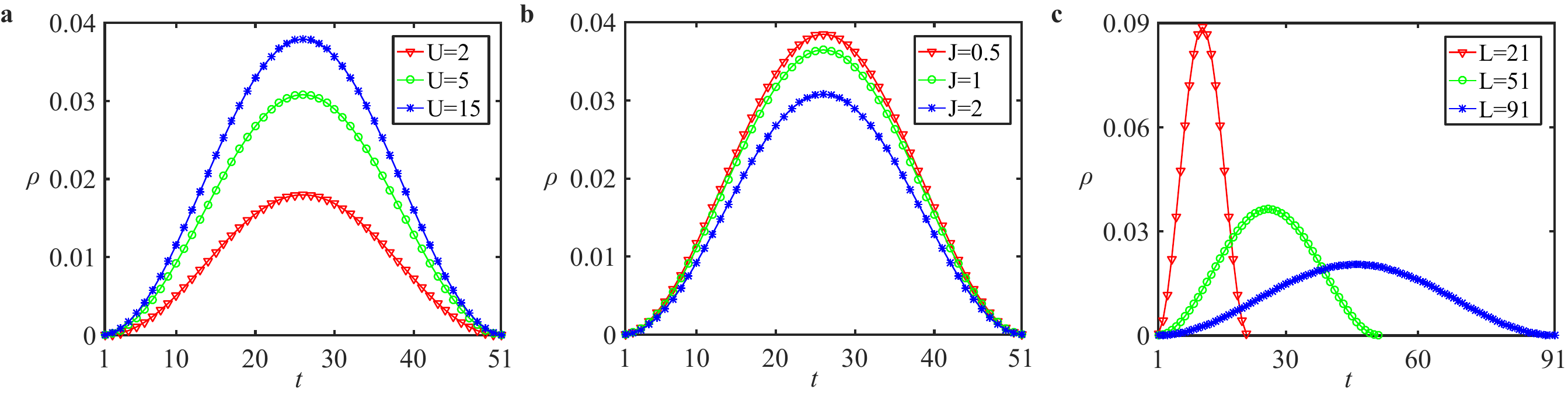}
\caption{\textbf{a-b} For a two-particle Hubbard model of lattice size $L=51$, localization given by $\mathcal{N}(R-r_c)$ can be analyzed by the density profile $\rho$ with different tunneling amplitude  and interaction strength, upon setting $w=v=t$. In \textbf{a}, we set $J=1$ and in \textbf{b}, it is fixed $U=10$. The maximum of $\mathcal{N}$ is observed with $t=\frac{L+1}{2}$. \textbf{c} By altering the  lattice size $L$, the localization is changed drastically, suggesting that $\mathcal{N}$ is size dependent. } \label{fig6}
\end{figure*}
\begin{figure*}[t]
\centering
\includegraphics[width=0.97\textwidth,height=0.36\textheight]{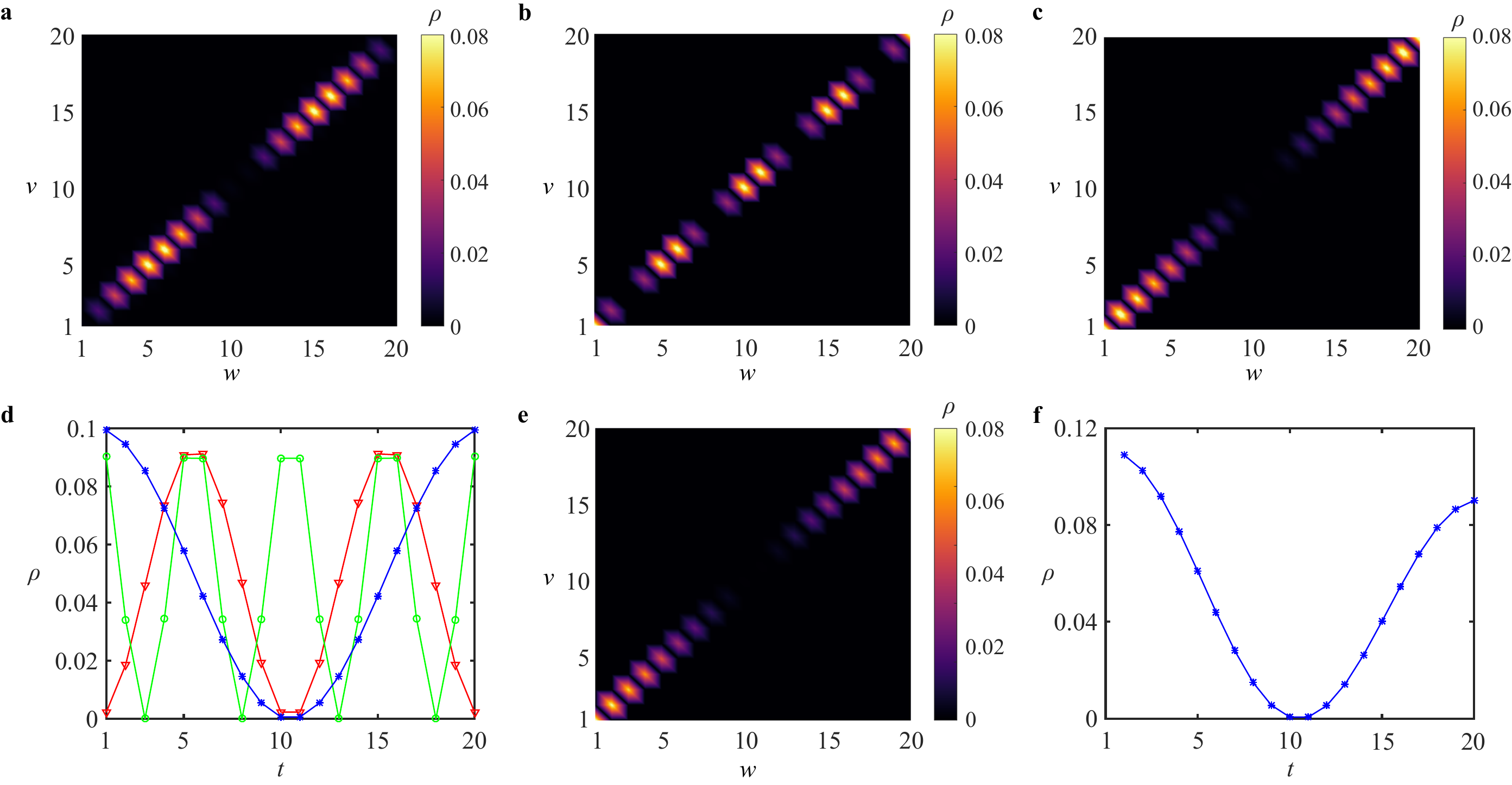}
\caption{For Hermitian two-particle Hubbard model with OBCs and $L=20,U=10,J=1$, the localization become diverse for bound state of different  eigen-energies, $E=10.75$ in \textbf{a}, $E=10.07$ in \textbf{b}  and $E=10.00$ in \textbf{c}.  \textbf{d} diverse $\mathcal{N}$ is obtained in correspondence to \textbf{a} red lines, \textbf{b} green lines, and \textbf{c} blue lines.  \textbf{e} Changing to non-Hermitian cases by setting $J^1_a=J^1_b=1.005, J^2_a=J^2_b=1.0$, U=10, localization of the bound state in \textbf{c} moves towards the left-down side and two maximums shrink to a single one. \textbf{f} In correspondence to \textbf{e}, density $\rho$ as a function of $w=v\equiv t$ becomes non-central symmetric.  } \label{fig6}
\end{figure*}
Our model shares the potential applications in the study of the non-Hermitian quantum optics.  By further including terms taking the form of two-photon tunneling, we present the topological nontrivial photon bound pairs, which are pinned at the edges of the diagonal line in the effective model. Meanwhile, considering the NHSE induced by the non-reciprocal tunneling, the doubly degenerate photon bound pair to be forced to the single left-down or the right-up corner.  To sum up, our works exhibit novel features of non-Hermitian many-body physics and will also benefit the future studies of the dissipative optical waveguide arrays.\\

\section*{Methods}
\subsection*{Localization described by $\mathcal{N}(R-r_c)$}
In Hermitian two-particle Hubbard model, the BSL localization in OBCs is indicated by $\rho^{H}\propto \mathcal{M}(r-r_0)\mathcal{N}(R-r_c)$, where $\mathcal{M}(r-r_0)$ suggests that two particles favor being bound to the same site and $\mathcal{N}(R-r_c)$ describes how the localization decays with the distance away from the lattice center. Characterizations of $\mathcal{N}(R-r_c)$ can be numerically achieved by setting $w=v\equiv t$, which is shown in Figure 8. It is observed that in contrast to  $\eta$ and $ \beta$, the localization scaling factor in $\mathcal{N}$ is size dependent. Besides, it should be noticed that the BSL is greatly related to the eigen-energies of bound states. For other (not the largest) eigen-energies, localization of the bound states can be diverse and multiple maximums are observed [see Figures 9a-9c]. Thus, it can be concluded that $\mathcal{N}(R-r_c)$  exhibits the dependence on energies. Before proceeding, we should mention that although the Hamiltonian in Eq.(1) shares the same non-reciprocal tunneling formalism  as the Hatano-Nelson model, the total localization of the two systems can be extremely different. Such features are endowed by the two-body interactions.

The competition between  $\mathcal{N}$ and  NHSE can be described through how the center of bound state is tuned towards the corner. Specifically, considering the cases $\beta_1<(>)0,\beta_2>(<)0$, the BSL is partially tuned towards the left-up (or the right-down) side and multiple maximums (for bound states not of the largest eigen-energy) can still be preserved. However, as we focus the conditions that $\beta_1>(<)0, \beta_2> (<)0$, BSL can moderately be coalesced with NHSE. In such circumstances, the multiple maximums will shrink to a single one due to the effects of NHSE. Correspondingly, density profile $\rho$ as a function of $w=v=t$ becomes non-central symmetric.

\subsubsection*{Data and Code Availability}
All data and code are available from
the corresponding author upon request.

\section*{ACKNOWLEDGEMENTS}
We thank Zhesen Yang for the helpful discussions. Huan-Yu Wang is supported by the start-up funding of Fuzhou University XRC-23079. Lin Wen acknowledges the funding from by the NSFC under Grant No. 12175027 and No. 11875010. Xiao-Fei Zhang is supported by the NSFC under Grants No. 12175129, No. 12475004 and Shaanxi Fundamental Science Research Project for Mathematics and Physics under Grant No. 22JSY034.

\section*{AUTHOR CONTRIBUTIONS}
Huan-Yu Wang conceived the study, analysed the results and wrote the manuscript.  Ji Li and Wu-Ming Liu helped with the discussions. Wen Lin and Xiao-Fei Zhang helped with the draft manuscript.

\section*{DECLARATION OF INTERESTS}
The authors declare no competing interests.



%

\end{document}